\documentclass[11pt]{article}
\usepackage{moriond,epsfig}
\usepackage{array,amssymb}
\def\jp{\rm J/\psi}

\def\be{\begin{equation}}
\def\ee{\end{equation}}
\def\bea{\begin{eqnarray}}
\def\eea{\end{eqnarray}}

\def\ttb{\mathrm{t\bar{t}}}
\def\bbb{\mathrm{b\bar{b}}}
\begin{document}

\begin{picture}(80,100)
\put(-40,-210){\epsfxsize220mm\epsfbox{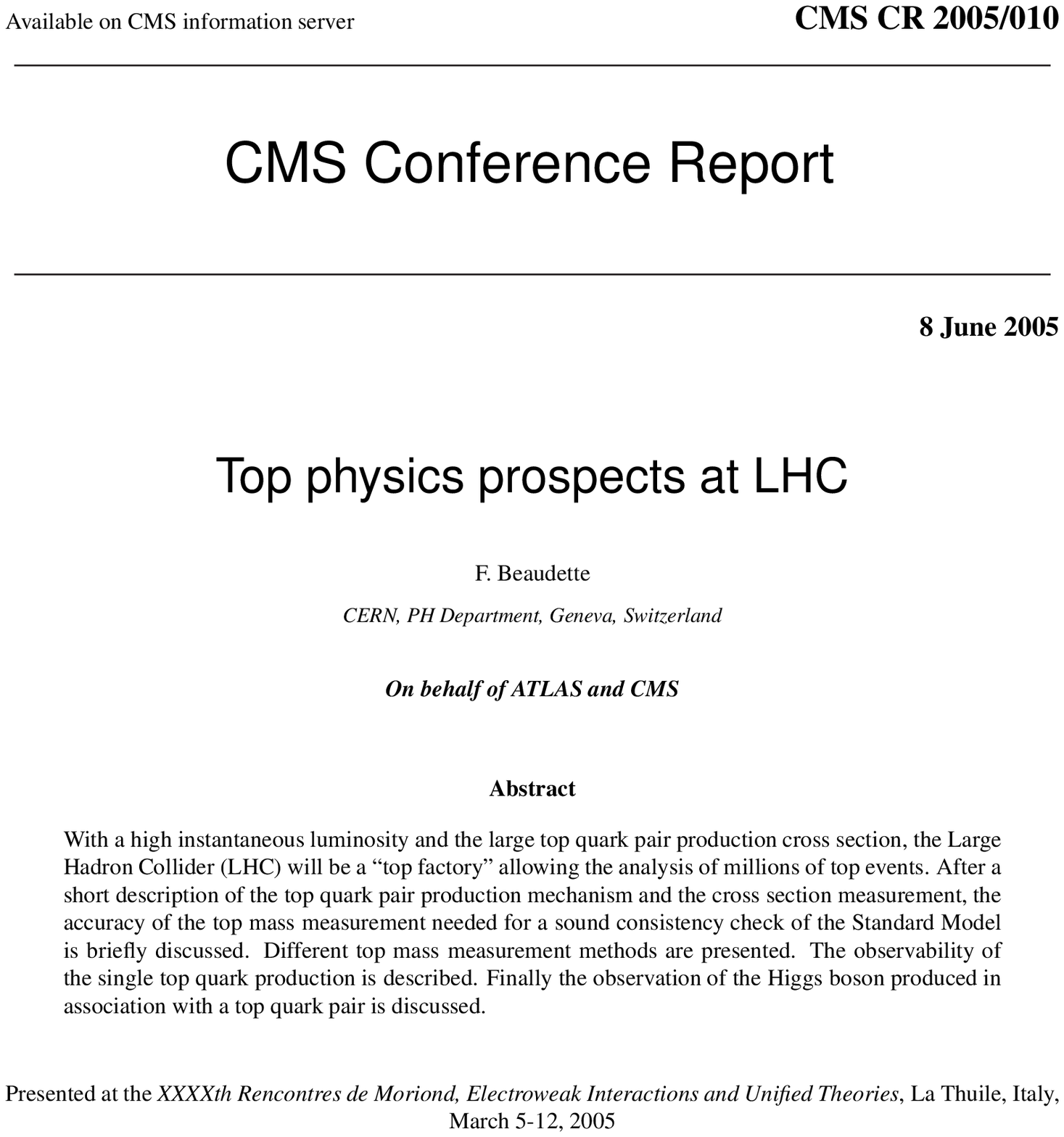}}
\end{picture}
\eject\null

\vspace*{4cm}
\title{Top physics prospects at LHC}

\author{ F. BEAUDETTE }

\address{CERN, PH Department CH-1211 Gen\`eve 23, Switzerland}

\maketitle\abstracts{
With a high instantaneous luminosity and the large top quark pair production cross section,
the Large Hadron Collider (LHC) will be a ``top factory'' allowing the analysis of millions of 
top events. 
After a short description of the top quark pair production mechanism and the cross 
section measurement, the accuracy of the top mass measurement needed for a sound consistency
check of the Standard Model is briefly discussed. Different top mass measurement methods are 
presented. The observability of the single top quark production is described. Finally the observation
of the Higgs boson produced in association with a top quark pair is discussed. 
}

\section{Top quark pair production}
\subsection{A top factory}
Whereas the centre-of-mass energy of the collisions at the LHC is seven times higher
than at the TeVatron, the  production cross section of top quark pairs is about hundred times
larger. It reaches~\cite{mangano} $\sigma_{\ttb} \approx 840$\,pb\,$(1\pm5\%_{\rm scale}\pm3\%_{\rm PDF})$ where the error
terms represent the systematic uncertainties related to the choice of the renormalization
and factorization scales and to the proton function structures respectively.
At $\sqrt{s}=14$\,TeV, the top quarks are mostly produced by gluon fusion (90\%). The 
quark annihilation, dominant at the TeVatron, amounts to only 10\% of the top
quark pair production. 

In the Standard Model (SM), the top quark always decays into a $\rm W$ and a $\rm b$ quark. As a result, 
the topology of the final state is mostly driven by the decay channels of the $\rm W$'s. 
The events where one of the $\rm W$'s decays into a lepton \footnote{Hereafter, ``lepton''
means electron or muon} have a clear signature: one isolated lepton, missing transverse energy
from the undetected neutrino and at least four jets of which two b jets.
At ``low'' luminosity, $L=10^{33}\,{\rm cm}^{-2}{\rm s}^{-1}$, there will be such so-called ``lepton+jet'' 
event every 4\,s while one top quark pair will be produced every second.
The LHC will thus be a ``top factory''.

\subsection{Top observation and cross section measurement}
\begin{figure}[h]
\centering
\includegraphics*[width=8cm]{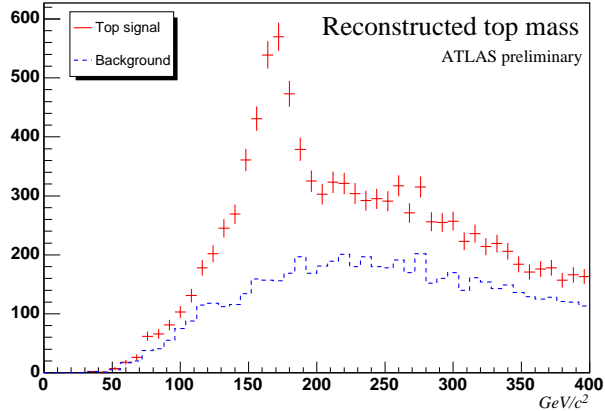}
\caption{Invariant mass of the three most energetic jets in the selected sample with 150\,pb$^{-1}$ of
integrated luminosity, as obtained from a fast simulation of the ATLAS detector.
The dashed blue line shows the W+4 jets background and the dots with error bars represent the
signal plus background expectations.}
\label{pallin}
\end{figure}

The production cross section is so large that the top signal will be visible after
the equivalent of one week of data taking at low luminosity~\cite{mangianotti} in the lepton+jet channel. 
By requiring one isolated lepton with a transverse momentum \mbox{$p_T> 20$\,GeV$/c$},
and exactly four jets with high transverse energy \mbox{($E_T>40$\,GeV)}, the top signal
is clearly visible above the $\rm W+4$ jets background in the  invariant mass distribution 
of the most energetic three jets (Fig.~\ref{pallin}).

The cross section can thus be measured. With the large number of events collected,
the statistical error will soon be negligible. After ``one month'' at low luminosity, it will
be at the level of 0.4\%. The overall error will be dominated by the systematic uncertainty related
to the luminosity measurement. A 5\% uncertainty is achievable~\cite{yellowbook}.
Because of its strong dependence on the top mass~\cite{mangano}, a measurement of the production 
cross section together with a precise measurement of the top mass will provide a test of QCD. 
Alternatively, within the SM, the cross section measurement provides a mass estimate, with
a potential accuracy of 3\,GeV/$c^{2}$ precision can be reached. 
A direct measurement can, however, be done with a better precision. 

\section{The top mass measurement}
\subsection{Why measuring (precisely) the top mass ?}
\begin{figure}[ht]
\centering
\includegraphics*[width=7.5cm]{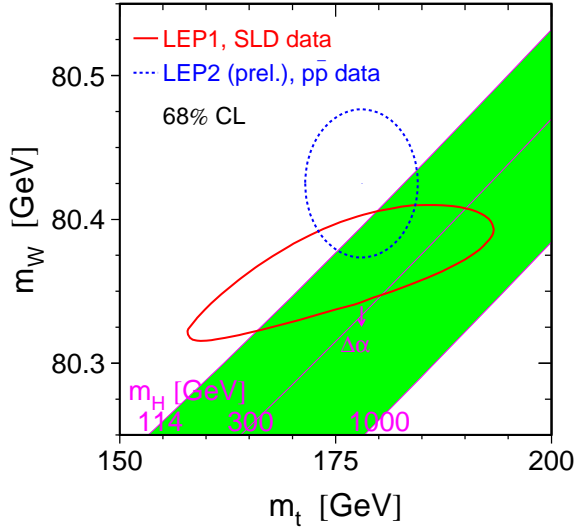}
\caption{Direct and indirect measurements of the $\rm W$ boson and top quark masses with lines of
constant Higgs boson masses in the Standard Model.}
\label{ew}
\end{figure}

Because of its mass, the top plays a particular r\^ole in the electroweak sector. 
In the SM, the W and Z boson masses are connected through the relation~\cite{yellowbook}
$m^2_{\rm W}(1-\frac{m^2_{\rm W}}{m^2_{\rm Z}}) = \frac{\pi\alpha}{\sqrt{2}G_\mu}\frac{1}{1-\Delta r}$,
where $G_\mu$ is the Fermi constant and $\Delta r$ contains the one-loop corrections. 
The top mass arises in $\Delta r$ via the loops in the W and Z boson propagators~\cite{willenbrock}
and gives rise to terms proportional to $m^2_{\rm t}/m^2_{\rm Z}$. 
Similarly, the Higgs boson loops give terms proportional to $\log{m_{\rm H}/m_{\rm Z}}$. 
The relationship thus obtained between the Higgs boson and top quark masses
is currently used as an indirect prediction of the Higgs boson mass~\cite{lepeww}: 
\mbox{$m_{\rm H}=126^{+73}_{-48}$\,GeV/$c^2$} for \mbox{$m_{\rm t} = 178 \pm 4$\,GeV/$c^2$}. 
The allowed region in the \mbox{($m_{\rm W}$,$m_{\rm t}$)} plane for different Higgs boson masses 
is displayed in Fig.~\ref{ew} as well as the direct and direct measurements of $m_{\rm W}$ and $m_{\rm t}$. 

At LHC, a direct measurement of the Higgs boson mass will be carried out towards a consistency
test of the SM by checking the relation between $m_{\rm t}$, $m_{\rm W}$ and $m_{\rm H}$.
To ensure a similar accuracy in the combination, the precision on $m_{\rm t}$ and $m_{\rm W}$ must 
fulfil~\cite{willenbrock2} $\Delta m_{\rm t} \approx 0.7\times 10^{-2} \Delta m_{\rm W}$ corresponding
to the slope of the constant Higgs boson mass lines in Fig.~\ref{ew}.
As can be seen in Table~\ref{table1}, a 2\,GeV/$c^2$ precision on $m_{\rm t}$ will allow a consistency
check of the SM with similar relevance as a 15\,MeV$/c^{2}$ accuracy on $m_{\rm W}$, reachable at LHC.
The LHC, however, can even do better and achieve a 1\,GeV/$c^2$ on $m_{\rm t}$ as described
in the following. A higher precision might
be needed in case of new physics discovery.
Such an accuracy would be obtained with an $\rm e^+ e^-$ linear collider~\cite{lincol}.
\begin{table}[htbp]
\caption{Expected precisions on the W boson and top quark masses at present and future colliders.} 
\vspace{0.4cm}
\begin{center}
\begin{tabular}{|c|c|c|c|}
\hline
Expected precision & $\Delta m_{\rm W}$ & $\Delta m_{\rm W}/0.7\times10^{-2}$  & $\Delta m_{\rm t}$  \\
& (MeV/$c^2$) & (GeV/$c^2$) & (GeV$/c^2$) \\
\hline
TeVatron & 25 & 4 & 3 \\
\hline
LHC & 15 & 2 & 1  \\
\hline
LC & 6 & 1 & 0.1\\
\hline
\end{tabular}
\label{table1}
\end{center}
\end{table}

\subsection{Measurement in the lepton+jet channel}
The lepton+jet channel is the golden channel for the top mass measurement. Indeed, 
the leptonic-decaying W boson allows the top events to be efficiently triggered and selected.
After the selection of the events with an 
energetic isolated lepton ($p_T>20$\,GeV$/c$) and a missing transverse energy in excess of 
20\,GeV,  the characteristics of the $\ttb$ events are then used to improve the purity of the sample.  
The events must contain at least four energetic jets ($E_T>20$\,GeV) of which two identified b jets.
The $\bbb$+jets, W+jets and Z+jets backgrounds are highly suppressed by this event selection~\cite{atlassummary}. 

The top mass is reconstructed from the two light jets from the W decay and the b jet from the top decay. 
As a result, the jet energy scale and angular resolutions are crucial.
The non b-jet pair minimizing the $(M_{\rm jj}-m_{\rm W})^2$ difference, where $M_{\rm jj}$ is the invariant 
mass of the two jets, is assumed to originate from the hadronically decaying W. 
A difference smaller than 20\,GeV$/c^2$ is required.
It is finally combined with the b jet giving the highest reconstructed top transverse momentum. 

The cone algorithm used to reconstruct the jets tends to underestimate the opening angle between the two jets 
from the W~\cite{roy}. An in-situ calibration can however be applied to correct the jet energies and directions. 

\begin{figure}[htbp]
\centering
\includegraphics*[width=7cm]{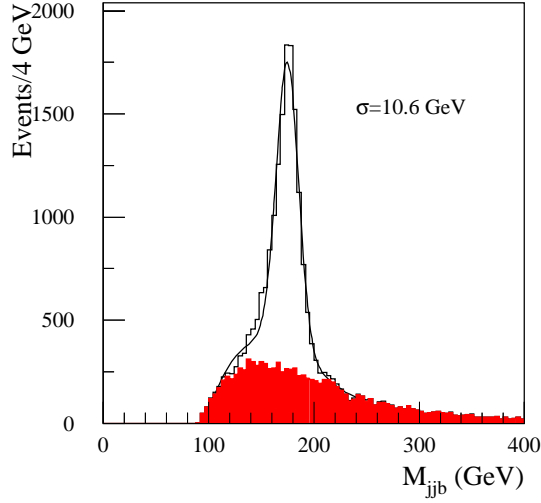}
\caption{Distribution of the jjb invariant mass of the selected events as obtained from a fast simulation
of the ATLAS detector for a 10\,fb$^{-1}$ integrated luminosity. The shaded area represents the combinatorial
background.}
\label{finaltopmass}
\end{figure}

The distribution of the three-jet invariant mass is displayed in Fig.~\ref{finaltopmass}. 
The reconstructed top quark mass is deduced from the fit value of the peak. 
The combinatorial background is dominant. With an integrated luminosity of \mbox{10\,fb$^{-1}$}, 
the statistical uncertainty on the top mass is at the level of \mbox{100\,MeV$/c^2$}.
The systematic uncertainties are summarized in Ref.~\cite{atlassummary}. 
The main two sources of systematic uncertainty are the final
 state radiation (FSR) and the b-jet energy scale. 
The FSR systematic error is conservatively evaluated as 20\% of the shift in the fit top mass when disabling the FSR 
and amounts to 1\,GeV$/c^2$.
At LHC, the light and b-jet energy scales are expected to be determined with
a 1\% precision~\cite{atlastdr}. In this analysis, the b-jet energy scale systematic uncertainty is
0.7\,GeV$/c^2$ whereas the light-jet energy scale uncertainty is mostly canceled by the in-situ calibration and 
amounts to 0.2\,GeV$/c^2$.
Altogether a 1.3\,GeV$/c^2$ error on the top mass is achievable. The effect of the FSR can be lowered down
to 0.5\,GeV$/c^2$ if a kinematic fit is implemented.
Indeed, the events with large FSR tend to have a high $\chi^2$ and can be removed from the analysis.
The systematic uncertainty thus becomes 
0.9\,GeV$/c^2$, dominated by the b-jet energy scale determination. 
As explained in the next section, it is possible to get rid of the heavy-flavour-jet-energy-scale related
uncertainty.

\subsection{Measurement in leptonic final state with $\jp$}
\begin{figure}[htbp]
\centering
\begin{tabular}
{>{\centering}m{7.5cm}
 >{\raggedright}m{5.5cm}
 }
\includegraphics*[width=7.3cm]{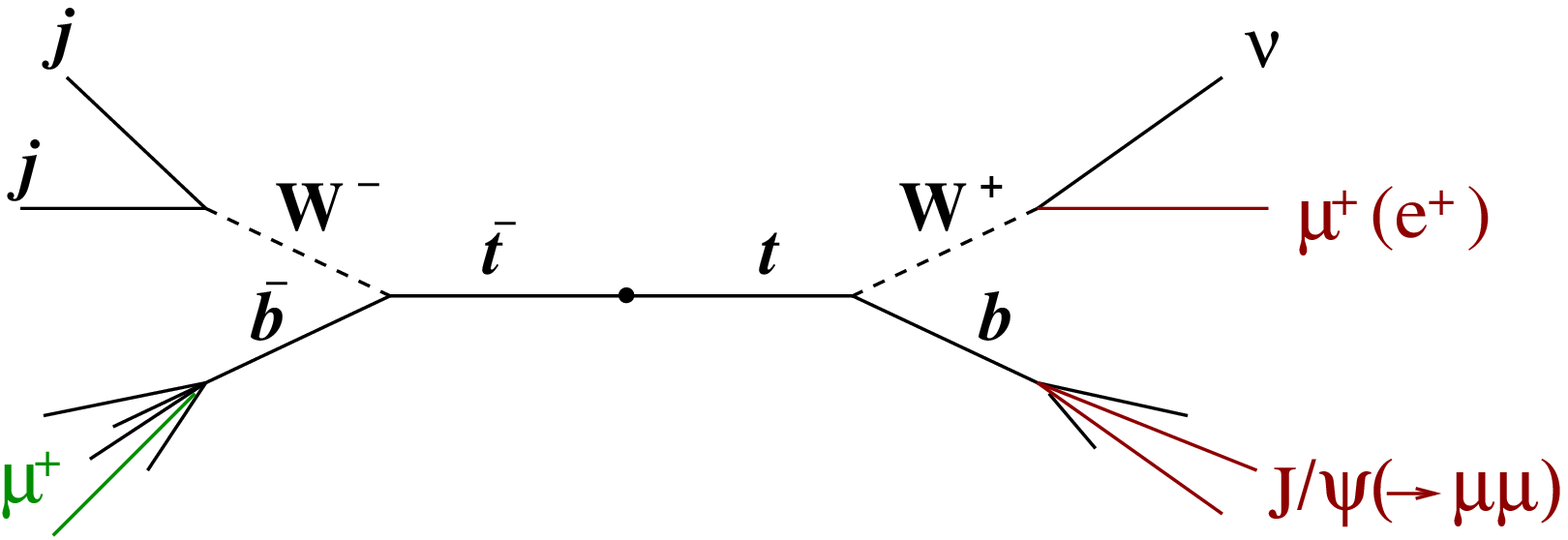} &
\includegraphics*[width=5.3cm]{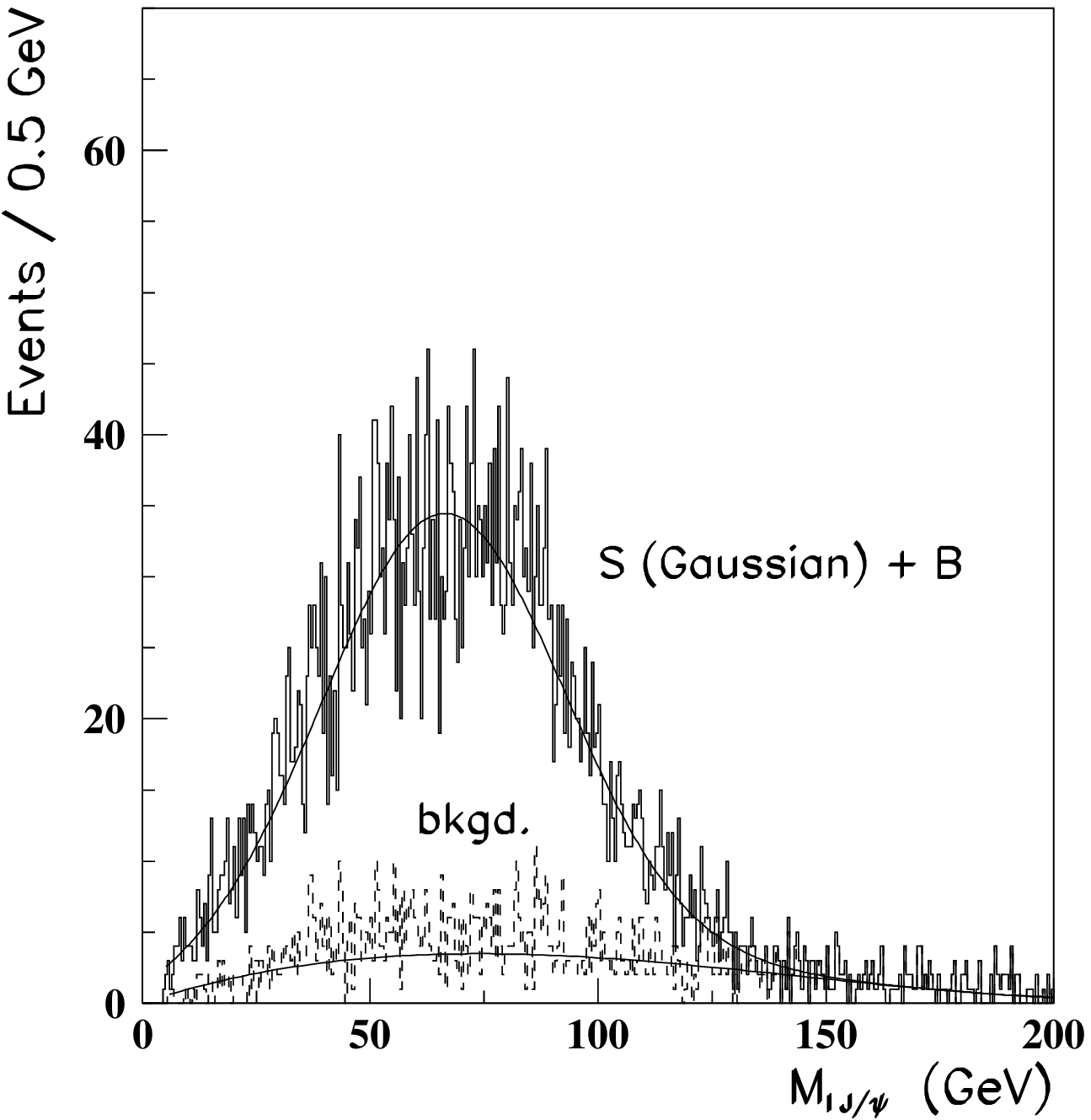} 
\end{tabular}
\caption{Diagram of the top decay to leptonic final state with $\jp$(left). 
Example of lepton-$\jp$ invariant mass in the four-lepton final state as obtained from
a fast simulation of the CMS detector after four years at high LHC luminosity (right).}
\label{jpsichannel}
\end{figure}

A determination of the top mass quark mass can be carried out in the lepton+jet
events where a $\jp$ arises from the b quark associated to the leptonic decaying W 
(Fig.~\ref{jpsichannel}). The top quark is partially reconstructed from the isolated
lepton coming from the W and corresponding b quark~\cite{cmsjpsi}. 

\begin{figure}[htbp]
\centering 
\includegraphics*[width=7cm]{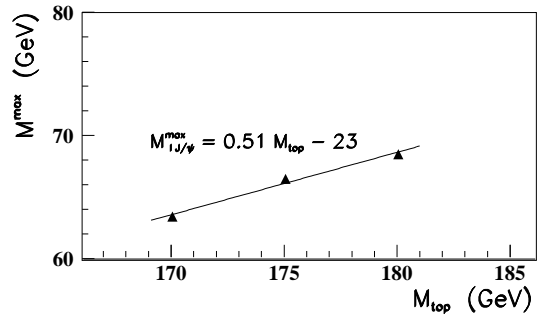}
\caption{Correlation between the fit value of the peak of the lepton-$\jp$ invariant mass distribution 
($\rm M^{\rm max}$) and the generated top mass as obtained from
a fast simulation of the CMS detector.}
\label{jpsichannel2}
\end{figure}

To solve the twofold
ambiguities on the b quark origin, a flavour identification, requiring a muon of the same electric charge
as the isolated lepton, is applied.
The $\jp$ can be precisely identified and reconstructed when it decays into a muon pair. 
As a result, one isolated lepton and three non isolated muons are required, two of
them being consistent with the $\jp$. This configuration is very seldom: one thousand events
per year will be collected at high luminosity ($L=10^{34}\,{\rm cm}^{-2}{\rm s}^{-1}$). 
The isolated lepton-$\jp$ invariant mass is determined (Fig.~\ref{jpsichannel}) and the
fit value of the peak turns out to depend linearly on the generated top 
mass~\cite{cmsjpsi} (Fig.~\ref{jpsichannel2}).

The background is essentially combinatorial, and its shape can be extracted from the data. 
The main systematic uncertainty comes from the b-quark fragmentation and is the combination
of the uncertainty on the b hadron spectrum in top decays and that of the $\jp$ spectrum in b hadron decays. 
The B factories can help in the determination of the latter.
An overall error on the top mass of the order of 1\,GeV$/c^2$ can be achieved. 

\section{Search for single top}
\begin{figure}[htbp]
\centering
\begin{tabular}{ccc}
\includegraphics*[width=6cm]{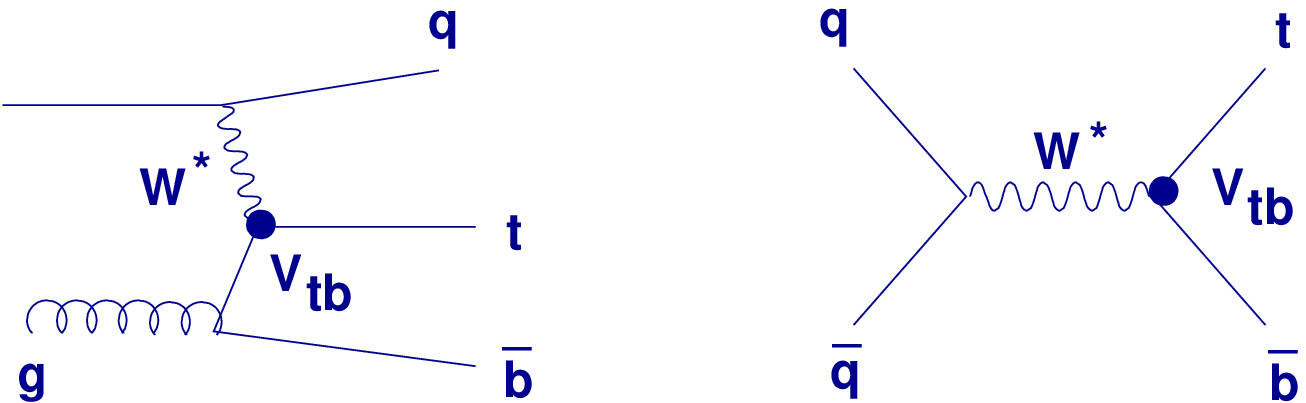} &
\mbox{~~~~~~~~~~} &
\includegraphics*[width=6cm]{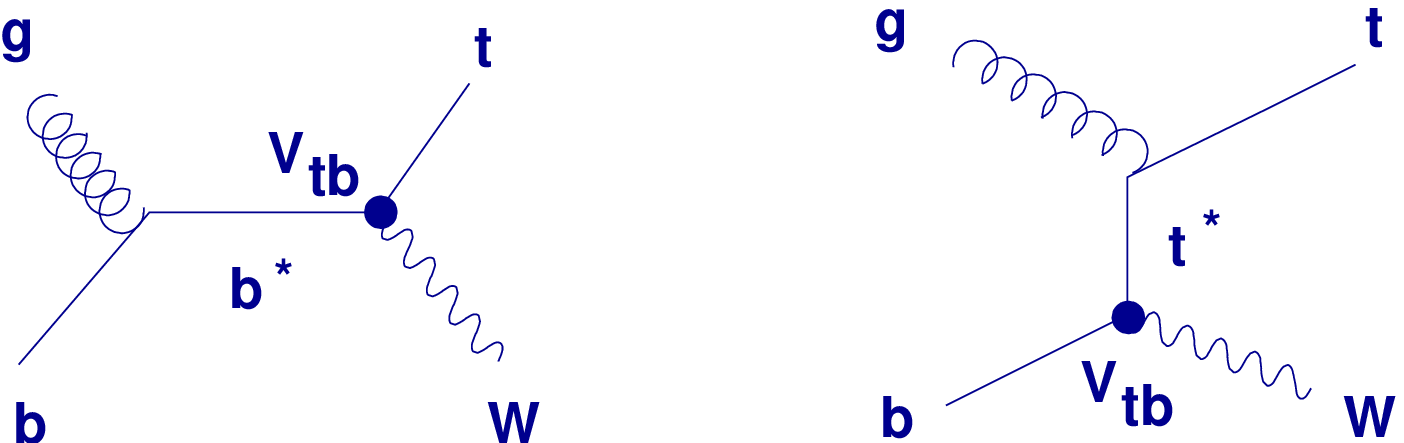} 
\end{tabular}
\caption{Electroweak top production diagrams in the $t$ and $s$ channels (left) and associate production
(right) }
\label{singletop2}
\end{figure}

The top quark can also be produced by electroweak interaction. In this case, one single top
is produced at a time. The total production cross section reaches is 310\,pb. The 
production diagrams at tree level are displayed in Fig.~\ref{singletop2}.
The dominant process is the W-gluon fusion \mbox{($t$ channel)} with a 
cross section of about 250\,pb. In the $t$ and $s$
channels, the production rate of top quarks is about 50\% higher than anti-tops~\cite{lukas} , while at the 
TeVatron, they are identical.
The associate production cross section is about 50\,pb and is one of the dominant backgrounds to the
search for the Higgs boson in the 
$\rm H\rightarrow WW^* \rightarrow$$\ell\nu \ell\nu$ channel~\cite{atlastdr}.

\begin{figure}[htbp]
\centering
\includegraphics*[width=7cm]{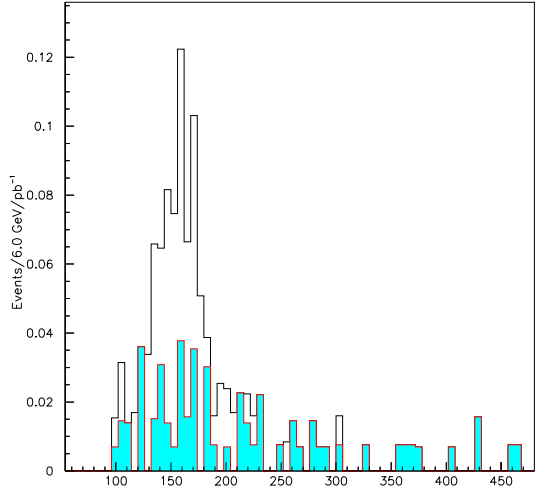}
\caption{Spectrum of the Wb invariant mass for the selected events obtained with a CMS full
 simulation. The open histogram represents the signal plus background expectations, the shaded 
histograms shows the background considered: $\ttb$, W+2 jets and W+3 jets.}
\label{finalsingletop}
\end{figure}
The event preselection, as in the lepton+jet channels, requires a leptonically decaying W. 
The different event topologies need dedicated final selections.
The $t$ channel is  taken here as an example. The full analysis is described in Ref~\cite{singletop}. 
The b jet from the initial gluon splitting is lost in the beam pipe. The events with
one forward non b-tagged jet and one central b-tagged jet (coming from the top) are selected.
The Wb invariant mass is then computed, the twofold ambiguity on the neutrino longitudinal momentum 
is solved by choosing the smallest one. This is true in only 55\% of the cases. 
The result is shown in Fig.~\ref{finalsingletop}. The overall efficiency (including the W to lepton
branching ratio) is 0.3\%. More than 6000 events are expected in 10\,fb$^{-1}$ of integrated 
luminosity. 
The main backgrounds are $\ttb$ and W+$\geq$2 jets. A signal-to-background ratio of 3.5 is obtained.

The single top production cross section can be measured with a 10\% precision, which is equivalent
to a 5\% precision on the measurement of the $\rm V_{tb}$(=1)  element of the CKM matrix.
The single-top polarization can also be measured in this 
channel with a 1.6\% statistical precision~\cite{yellowbook} with 10\,fb$^{-1}$.

\section{Associate Higgs boson production}
\begin{figure}[htbp]
\centering
\includegraphics*[width=8cm]{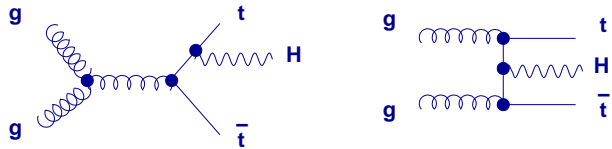}
\caption{Associate $\ttb\rm H$ production diagrams}
\label{ttbh1}
\end{figure}

For small Higgs boson masses ($\lesssim$ 130\,GeV$/c^2$),
the $\rm H\rightarrow \bbb$ decay channel is dominant. Unfortunately,
it is impossible to efficiently trigger the acquisition of these events due to the huge
di-jet $\bbb$ background present at LHC.
To observe the $\bbb$ decay of the Higgs boson, an associate production mode (with W, Z bosons or with 
a $\ttb$ pair)  has to be considered.
The $\ttb\rm H$ production diagrams are 
presented in Fig.~\ref{ttbh1}. These channels allow the top Yukawa coupling to be measured.
The cross section is  small: $\sigma(m_{\rm H}=120$\,GeV$/c^2)=0.8$\,pb, while 
the $\ttb\bbb$ background has a 3\,pb cross section.

The lepton+jet events are first selected. The final state is intricate, since in addition of the ``usual''
lepton+jet event, two additional b jets from the Higgs boson are present. As a result,
the event selection requires at least six jets in the final state of which exactly four b jets. 

\begin{figure}[htbp]
\centering
\begin{tabular}{cc}
\includegraphics*[width=7cm]{ttbHCMS.epsi} &
\includegraphics*[width=6.6cm]{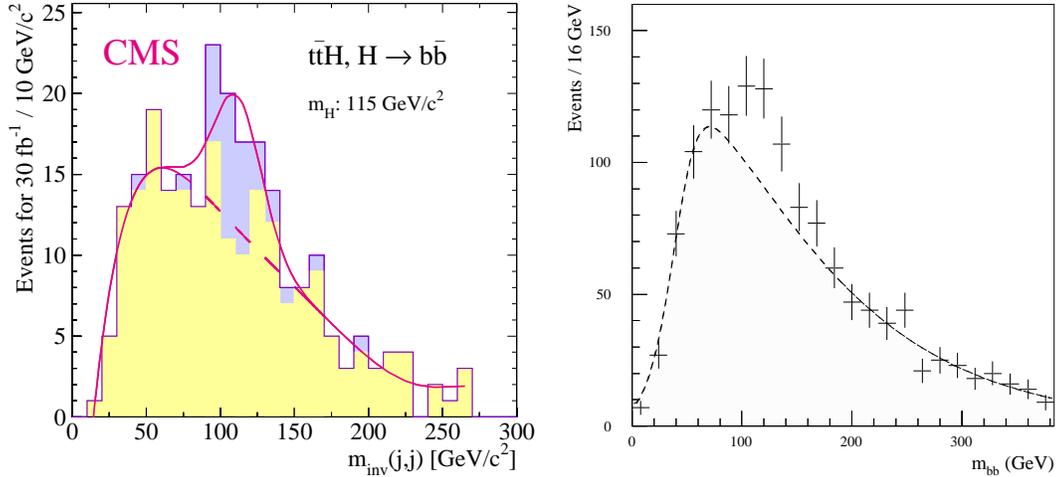} 
\end{tabular}
\caption{Signal plus background expectation for the Higgs boson reconstructed mass in the
 $\ttb\rm H$ channel with $\rm H\rightarrow \bbb$
in the CMS and ATLAS detectors for
$m_H=$115\,GeV$/c^2$ and $m_H=$120\,GeV$/c^2$ with 30\,fb$^{-1}$ and 100\,fb$^{-1}$ respectively. 
In both cases, fast simulations of the detectors have been used.}
\label{ttbh}
\end{figure}

Both W's are fully reconstructed. The two b's from the top decays
have to be identified and the pair giving the ``best'' reconstructed top quarks pair 
is chosen. 
The remaining two b's are combined to reconstruct the Higgs boson. The resulting
invariant mass distribution is shown in Fig.~\ref{ttbh}, showing a nice agreement between the ATLAS
and CMS analyses and a peak due to the presence of the Higgs boson. 

The shape of the background can be extracted from $\ttb\rm jj$ data. 
With 30\,fb$^{-1}$, 40 signal events are expected~\cite{yellowbook},
with a significance of 3.6\,$\sigma$. A 16\% precision on the Yukawa 
coupling should be reached. 
The combination of the low and high luminosity runs giving a  integrated luminosity of 100\,fb$^{-1}$
will allow a 4.8\,$\sigma$ significance and
a 12\% precision on the Yukawa coupling to be reached. All these numbers are relative to a Higgs boson mass of
120\,GeV$/c^2$.

\section*{Conclusion}
The physics of the top quark will be one of the LHC main topics. 
Many exciting analyses will be carried out. Only a few of them have been summarized in this paper.
Most of the analyses can be done with the the first 10\,fb$^{-1}$. Due to the large
production cross section, the statistical uncertainty will be, in most of the cases, quickly negligible. 

The top mass measurement will be a key issue. A \mbox{1\,GeV$/c^2$} precision can be reached provided that an excellent
understanding of the detectors to control the systematic uncertainties. 
The study of the top quark sector  highlights several theoretical challenges like the high order 
QCD calculations and the b fragmentation. 
Finally, most of the analyses presented in this report 
make use of fast simulation of the ATLAS and CMS detectors. As a result, the
systematic studies are ahead of us.

\section*{Acknowledgments}
I would like to thank the conference organizing committee for their hospitality  and 
financial support.

\end{document}